\documentclass[fleqn,10pt]{wlscirep}
\title{Navigability of temporal networks in hyperbolic space}

\author[1, 2]{Elisenda Ortiz}
\author[1, 2]{Michele Starnini}
\author[1, 2, 3,*]{M.Ángeles Serrano}
\affil[1]{Departament de F\'{\i}sica de la Mat\`eria Condensada, Universitat de Barcelona, Mart\'{\i} i Franqu\`es 1, 08028 Barcelona, Spain}
\affil[2]{Universitat de Barcelona Institute of Complex Systems (UBICS), Universitat de Barcelona, Barcelona, Spain}
\affil[3]{ICREA, Pg. Llu\'{\i}s Companys 23, E-08010 Barcelona, Spain}

\affil[*]{marian.serrano@ub.edu}

\begin{abstract}
Information routing is one of the main tasks in many complex networks with a communication function. Maps produced by embedding the networks in hyperbolic space can assist this task enabling the implementation of efficient navigation strategies. However, only static maps have been considered so far, while navigation in more realistic situations, where the network structure may vary in time, remain largely unexplored. Here, we analyze the navigability of real networks by using greedy routing in hyperbolic space, where the nodes are subject to a stochastic activation-inactivation dynamics. We find that such dynamics enhances navigability with respect to the static case. Interestingly, there exists an optimal intermediate activation value, which ensures the best trade-off between the increase in the number of successful paths and a limited growth of their length. Contrary to expectations, the enhanced navigability is robust even when the most connected nodes inactivate with very high probability. Finally, our results indicate that some real networks are ultranavigable and remain highly navigable even if the network structure is extremely unsteady. These findings have important implications for the design and evaluation of efficient routing protocols that account for the temporal nature of real complex networks.
\end{abstract}
\begin{document}

\flushbottom
\maketitle
%
%
\thispagestyle{empty}

\section*{Introduction}

Transfer of information, mass, or energy is a key function in many natural and artificial complex systems, 
ranging from gene-regulatory networks~\cite{citeulike:3149352} and the brain~\cite{Papo2014} to online and offline social networks~\cite{DBLP:conf/sigcomm/Zhang13}, the Internet~\cite{marian_greedy}, and transportation networks~\cite{Bast:2010:FRV:1888935.1888969}.  Milgram's experiment~\cite{Travers69anexperimental} showed that some of these systems can be efficiently navigated, 
i.e., their elements are able to perform an effective information routing even though they do not possess global knowledge of the system. 
This surprising property was first explained by Kleinberg using a network model~\cite{kleinbergsearch,Kleinberg:2010}, 
in which each node resides in the Euclidean plane and forwards information to the neighbor which is closer to destination. 
More recently, it has been suggested that the geometry of complex networks is not Euclidean but hyperbolic, as a result of the interplay between the popularity and similarity attributes of the nodes \cite{Serrano:2008ga,PhysRevE.80.035101,Boguna:2010aa,papadopoulos2012popularity}. 
Within this framework, the observed topological properties of complex networks are naturally explained on the basis of a hidden metric space defining distances between nodes, and a connection probability dependent on such distances. 
Moreover, distances in the underlying hyperbolic geometry can guide greedy routing very efficiently in scale-free networks, 
meaning that the success probability of the process is extremely high, while the routing paths deviate only slightly from the topological shortest paths, following closely the geodesics in the hyperbolic plane \cite{Boguna2009}.

 These advances in the understanding of the navigability of complex networks are framed 
 within the traditional approach taking the structure of networks as static. 
 However, this assumption has been recently challenged by the empirical observation of a temporal dimension in many natural and social systems 
\cite{PhysRevLett.92.108501, Thompson096461, 0295-5075-81-4-48002, journals/corr/abs-1106-0288},
demonstrating that nodes and edges switch on and off with several time scales.
The empirical analysis of such temporal networks \cite{temporalnetworksbook}
has unveiled new statistical properties, 
such as a heavy-tailed distribution of inter-event times between consecutive links, known as burstiness \cite{Barabasi:2005uq},
or the heterogeneous distribution of activity in social interactions \cite{Perra:2012aa}.
Temporal effects have been shown to impact both
the behavior of dynamical processes on networks 
\cite{PhysRevE.83.025102, dynnetkaski2011, PhysRevE.85.056115, burstylambiotte2013,GarciaPerez:2015} and the connectivity of their corresponding static representations \cite{Parshani:2010, PhysRevE.94.022316}. Time-respecting paths \cite{temporalnetworksbook}, for instance, play a crucial role in slowing down or speeding up the spreading of information or diseases \cite {Scholtes:2014aa}, and certainly affect also the message routing throughout the network. 

Although navigation is expected to be substantially different in temporal networks than in static ones, 
few empirical or theoretical works have been devoted to study the impact of the temporal dimension on the navigability of complex systems \cite{marian_greedy, George:2007:SND:1784462.1784488, Kirst:2016aa}. 
Some of these studies are concerned with the small world property~\cite{Tang2010}, while others aim at quantifying network vulnerability to temporary failures~\cite{Trajanovski2012}, or explore temporal networks using greedy walks that proceed from node to node by always following the first available contact~\cite{Saramaki2015}. 
However, the general mechanisms that guarantee an optimal routing in situations 
where the network's structure changes with time, 
or where noise affects the communication paths, are not fully understood yet. 
Uncovering such mechanisms is thus a fundamental task, with a broad range of potential applications, 
for instance, in communication engineering \cite{Naumov:2005:SRM:1120164.1120179} and system biology \cite{Tononi:aa}. 

Here, we tackle this issue by proposing a hybrid model to study the navigability of temporal networks and show that, surprisingly, temporal networks can be navigated more efficiently than their static counterparts.
Furthermore, we show that some real networks are ultranavigable, meaning that they remain highly navigable even when the network topology is strongly dynamic.
Our model considers static reconstructions of real networks and a simple node activation-inactivation dynamics. 
 This allows us to control for the maximum duration of the routing process, as well as to discard peculiar features of specific real evolving systems, such as circadian rhythm \cite{Jo_NJP2012}. The activation dynamics may represent temporal failures of nodes due to random unknown events, or noise. 
Our approach suggests a new greedy routing protocol in static networks, that combines standard greedy routing and a simulated activation dynamics, which can boost the navigability of some real networks,  at the expense of elongated paths.

Next, we set our analysis upon five different empirical networks: ArXiv collaborations (ArXiv), US Commodities networks (Commodities), Metabolic networks (Metabolic), the Internet at the autonomous system level (Internet) and the World Trade Web (WTW). Detailed descriptions of the data sets can be found in Methods.

\section{\label{sec:level1}Greedy routing on temporal networks}

Information packets, or other assets, are transferred in a network from a source node to a destination one by following greedy routing
in hyperbolic space \cite{Boguna:2010aa}. We consider a two-dimensional hyperbolic plane of constant negative curvature where each node $i$ has polar coordinates ($r_i$, $\theta_i$), see Methods. 
The implementation of the routing algorithm requires that there is only one packet per source-destination pair, that each node knows its coordinates, the coordinates of its neighbors in the network, and the coordinates of the destination node. 
Then, the node holding the packet will transfer it to its neighbor with the smallest hyperbolic distance to the destination node. 

We take the hyperbolic embedding of the largest connected component~\cite{Newman2010} of each real complex network, 
that we refer as the static map $\mathcal{M}(G_0,S)$, where $G_0$ stands for the static graph and $S$ is the underlying metric space where the nodes have permanent coordinates. Next, we generate several synthetic temporal networks by applying a Poissonian activation-inactivation dynamics on its nodes. 
We consider that nodes can be in an active state, being able to receive and forward information, or in an inactive state, in which case they cannot receive neither forward information packets.
At each time step $t$, each node $i$ is active with probability $a_i$. 
Thus, at each time step $t$, a graph $G_t$ is defined, in which only active nodes and the links between them are present.  
The sequence of graphs $\mathcal{G}=\{G_t\}_{t=1,2,...T}$ constitutes a synthetic temporal network of length (duration) $T$.
The activation probabilities control the density of the temporal networks, 
affecting the probability of a message being sent.
For instance, in the case of a constant activation probability set equal for all nodes, $a_i = a$, 
each graph $G_t$ has an expected average degree equal to $\overline{k}_t = a \ \overline{k}$, 
where $\overline{k}$ is the average degree of the original static network.

Therefore, the greedy routing acts on a temporal map $\mathcal{M}(\mathcal{G},S)$ depending on the temporal network $\mathcal{G}$ and the underlying hyperbolic space $S$.
The greedy forwarding algorithm is implemented sequentially on the temporal map $\mathcal{M}(\mathcal{G},S)$, so there is one attempt to forward the information packet for each time step $t$. At time $t$, the node holding the information packet tries to forward it to its neighbor with the lowest distance to the final destination. If the neighbor is active at time $t$, then it receives the packet. Otherwise, the packet remains at the holding node.
The model with $a=1$ corresponds to greedy routing on the original static network, with all nodes active at all times, 
for a number of steps equal to $T$. 
Therefore, the network's duration $T$ can be interpreted as the maximum lifetime of information packets. 
In this scenario, a greedy path is successful when a packet reaches its destination in a time $t \le T$,  and unsuccessful otherwise. In the limit of $T\rightarrow  \infty $, all packets are expected to be able to reach their destination because the number of different paths that can be realized by greedy routing on the temporal networks grows with $T$.
 
We run numerical simulations for different network's duration $T$, taking a number of random source--destination pairs which is the minimum between $10^5$ and $N(N-1)/2$, where $N$ is the number of nodes of the network. In numerical experiments varying the activation probability, the random subset of source--destination node pairs is kept the same, while it is changed when varying $T$.

\begin{figure*}[th]
\includegraphics[width=1.0\textwidth]{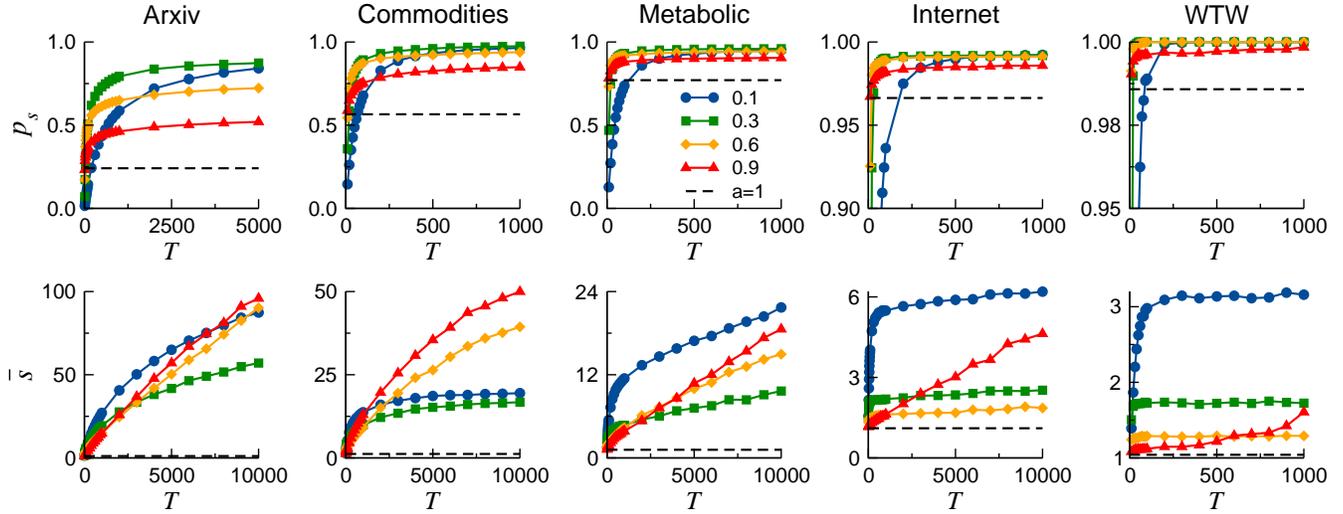}
\caption{\textbf{Success ratio $p_s$ (top row) and average stretch $\overline{s}$ (bottom row) as a function of $T$}, for different values of the activation probability $a$, in five real networks. The success ratio and the average stretch in the static map, $a=1$, appear plotted with dashed lines.
\label{fig:success}}
\end{figure*}

\section{\label{sec:level1}Results}

We first consider a constant activation probability set equal for all nodes, $a_i = a$.  In this case, the model is characterized by two parameters, the activation probability $a$, which controls the activation dynamics, and the network's duration $T$, which represents the maximum lifetime of information packets.
  We evaluate the performance of greedy routing on the temporal map by measuring two main quantities:
 the \textit{success ratio} $p_s$, defined as the fraction of packets that successfully reach their destination within a time $T$ over the total number of source-destination pairs considered; and the \textit{average topological stretch} $\bar{s}$ of successful greedy paths, where the stretch is defined as the ratio between the hop-length of a greedy path and the shortest path between the corresponding source and destination nodes. The stretch tells us how much the successful greedy paths are longer with respect to the shortest ones.

In the Supplementary Material, we also give results for the average geometric stretch $\overline{s}_g$, which is defined analogously to $\overline{s}$ but considering the hyperbolic lengths of greedy and shortest paths; and the average coverage $\overline{\kappa}$, which informs of the average number of different visited nodes against the average number of nodes that compound a successful path.
   
\subsection{\label{sec:level2_Effects}Effects of network dynamics on navigability}

The success ratio $p_s$ is a key parameter in determining the navigability of complex networks. 
A large success ratio, close to $p_s \sim 1$, means that almost all nodes can be reached by a message sent by any other node. 
On the contrary, if $p_s$ is small, information can not be successfully transmitted from most nodes.
Fig.~\ref{fig:success}, top row, shows the fraction of successful paths $p_s$ as a function of the network duration $T$, for different values of the activation probability $a$.
The success ratio varies considerably across different static networks ($a=1$), ranging from very low success ratio for the ArXiv, to $p_s \sim 1$ for the WTW and the Internet, which indicates a better congruence of these systems with their underlying geometry. Remarkably, for sufficiently large $T$, the success ratio in all temporal networks ($a<1$) under consideration is always larger than the one achieved on their static counterparts (dashed line, top row Fig.~\ref{fig:success} and Table~\ref{tab:1} in Methods). 
 This effect is particularly evident for the cases where $p_s$ in the static map is rather low, such as for the ArXiv network, where the success ratio increases from $p_s = 0.24$ for $a=1$ to $p_s= 0.90$ for $a\sim 0.2$. Nonetheless, when the static success ratio is high (e.g. Internet), $p_s$ on the temporal maps increases too.
 
As expected, $p_s$ is a growing function of the network duration $T$:
the larger the maximum lifetime of the packets, the higher $p_s$. In the limit of $T \rightarrow \infty$, $p_s$ is expected to reach its maximum since, for any pair of nodes, 
all different paths between them will be available at some time, 
ensuring that a successful one will certainly arise. 
This implies that the success ratio always increases with $T$, although the growing rate can be extremely slow for very large $T$. 
Oppositely, in the routing on static networks, $p_s$ does not vary with $T$ because no new paths are added by increasing the lifetime of information packets.

Our results show that, surprisingly, it is more efficient to have some (or even a great number of) nodes inactive than having all nodes active and contributing to the routing process. The reason for this behavior is rooted in the fact that, with $a=1$, some packets might get stuck into \textit{topological traps}. 
From the greedy routing definition, indeed, it is clear that if a packet comes back to a node twice, it will come back again, and the loop would continue forever with the packet never reaching its destination. 

\begin{figure}[tbp]
\centering
\includegraphics[width=0.42\textwidth]{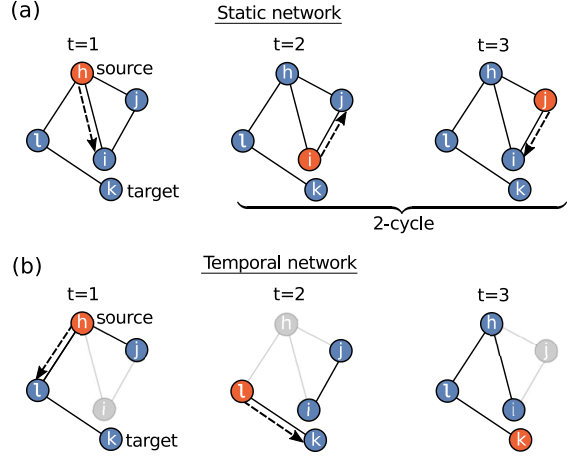}
\caption{\textbf{Representation of a topological trap,} in Euclidean space. Greedy routing demands sending the information packet always to the neighbor closest to destination. Consequently, in \textbf{a)} the packet never stops jumping between nodes $i$ and $j$ and is captured by the topological trap,  while in \textbf{b)} the inactivation of $i$ enables the packet to follow an alternative route through $l$ and successfully reach destination node $k$.\label{fig:sketch}}
\end{figure} 

To understand this mechanism, consider
a node $i$ sending a packet to his neighbor $j$, because $j$  is the closest (among all $i$’s neighbors) to destination node $k$. If during the next time step, node $i$ turns out to be the closest node to destination $k$ among 
$j$’s neighbors, then the packet will return to $i$. As long as no topological change takes place in the
network, this process will repeat endlessly. Any cycle involving a packet coming back to a node twice constitutes
a topological trap, See Fig.~\ref{fig:sketch}. In contrast, if $a < 1$, the topology of
the network changes at each time step, hence the packet
is able to escape any topological traps it may encounter
along the route and eventually reach its destination. Nevertheless, the new successful path followed by the packet will deviate from the geodesic connecting the two nodes in the hyperbolic plane, thus the path length will necessarily be longer than the shortest.

The average topological stretch $\overline{s}$, 
defined as the ratio between the hop-length of greedy paths and the corresponding shortest paths in the network, is also a measure of navigation efficiency. 
From its definition, it holds that $\overline{s} \geq 1$. A small stretch,  $\overline{s} \gtrsim 1$,  indicates that most packets follow a route very close to the shortest one, while if $\overline{s} \gg 1$, paths are much longer. 
Fig.~\ref{fig:success}, bottom row, shows the average stretch $\overline{s}$ as a function of the network's duration $T$, for different values of the activation probability $a$.
As for the success ratio, $\overline{s}$ is also an increasing function of $T$. 
Indeed, the larger the duration $T$, the lengthier the paths that become successful, and these very long paths increase the average stretch. 

It is important to note that the shortest paths between two nodes in $\mathcal{M}(\mathcal{G,S})$ may be much longer than the shortest path in the corresponding static maps, because of time-respecting paths. 
This is particularly true for very sparse temporal networks, i.e. with low activation probability. 
Therefore, $\overline{s}$ is always greater in temporal maps than in the corresponding static ones, as shown in Fig.~\ref{fig:success}, demonstrating that the activation dynamics is responsible for creating lengthier successful paths. 

This effect is clearly visible in those data sets where $p_s$ in the static networks is low, such as the ArXiv or Commodities. 
In these networks, the large increase in the success ratio due to the activation dynamics comes with a large growth in the average stretch. 
The probability of finding much more successful tracks is increased at the cost of choosing longer paths. 
On the contrary, if $p_s$ in the static maps is high, such as for the Internet or the WTW, $\overline{s}$ shows a small increase in the temporal maps. 
These different profiles correspond to the different geometricity of the considered networks. In fact, the less congruent topology and geometry are, the larger the number of topological traps present in $\mathcal{M}(\mathcal{G},S)$ and the larger the potential increase in success. Temporal maps with limited congruency, such as ArXiv or Commodities, show the larger gains in success overcoming traps at the expenses of a notable increase in $\overline{s}$, and longer durations $T$. On the contrary, networks with a conspicuous latent geometry, like the Internet and the WTW, are not characterized by a large number of topological traps, hence $\overline{s}$ does not rise as much.

Interestingly, different effects are obtained on the success ratio and the stretch depending on the activation probability. 
The lowest values of $\overline{s}$ are found for intermediate values of $a$, while $p_s$ generally increases as the activation probability decreases, down to a value for which the network becomes too inactive, and then $p_s$ becomes lower again. 
In most networks, $p_s$ remains almost unchanged if the activation probability is set equal to 
$a=0.3$ or to $a=0.1$ in the limit of large $T$, 
while $\overline{s}$ significantly increases if the activation decreases from $a=0.3$ to $a=0.1$, specially for the Internet and the WTW.
Conversely, choosing $a=0.6$, $p_s$ grows from 0.96 in the static case to 0.99 for the Internet and from 0.99 to 0.997 in the WTW, 
but the stretch increases very little from $\overline{s} = 1.11$ to $\overline{s} \approx 1.76$ for the Internet, and from $\overline{s} =1.04$ to $\overline{s} \approx 1.29$ for the WTW. This indicates it may exist an optimal activation probability that maximizes the increase in the success ratio and minimizes the increase in the stretch.

\begin{figure*}[tb]
    \includegraphics[width=1.0\textwidth]{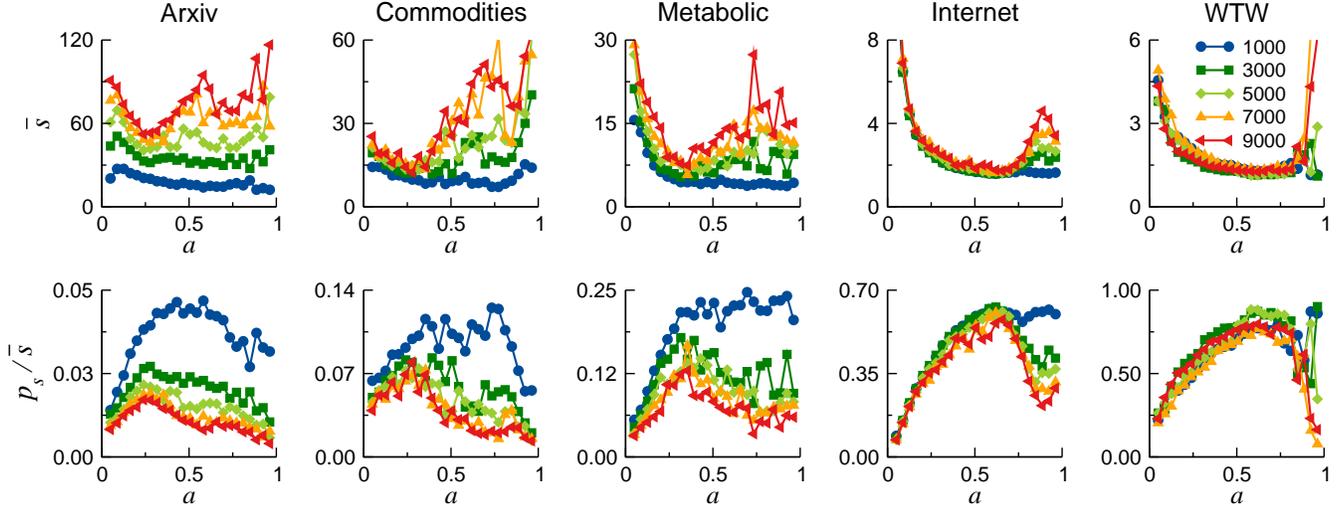}
  \caption{ \textbf{Average stretch, $\overline{s}$ (top row), and  success ratio divided by the average stretch, $p_s / \overline{s}$ (bottom row) as a function of $a$}. Each curve corresponds to a different value of the duration $T$. Notice the rightmost point is not 1 (the static reference) but 0.96. The optimal activation $a_O$ of each network is observed as a maximum in the plots of the bottom row. The approximate values of $a_O$ for each network are: $a_{O}^{\mathrm{Arx}}\approx0.25$, $a_{O}^{\mathrm{Com}}\approx0.25$, $a_{O}^{\mathrm{Met}}\approx0.33$, $a_{O}^{\mathrm{Int}}\approx0.60$ and $a_{O}^{\mathrm{WTW}}\approx0.67$.
        \label{fig:stretch_a}}
\end{figure*}

\subsection{\label{sec:level2}Optimal activation probability}

Top row of Fig.~\ref{fig:stretch_a} shows $\overline{s}$ as a function of the activation probability $a$, for several values of the network's duration $T$. 
Interestingly, the average stretch is not a strictly decreasing function of $a$, but it reaches a minimum for some intermediate value. 
On one hand, when the activation probability is very small, the stretch is typically large because of the lack of available active neighbors. The packet will usually remain in the holding node or it will be transferred erratically, resulting in an increase of $\overline{s}$. 
On the other hand, if $a \lesssim 1$, the topology of $\mathcal{M}(\mathcal{G},S)$ is similar to the static one, so the packet tends to fall into the same topological traps spending a long time moving in cycles (thus increasing the stretch) before it succeeds to escape the loop.
Remarkably, the minimum of $\overline{s}$ is reached for some optimal value of the activation probability, $a = a_{O}$.

This feature is addressed in more detail in the bottom row of Fig.~\ref{fig:stretch_a}, which shows the ratio between the success ratio and the average stretch ($p_s / \overline{s}$) as a function of the activation probability $a$.
Since $p_s \leqslant 1$ and $\overline{s} \geqslant1$ by definition, perfect navigability is reached when $p_s = \overline{s} = 1$, and thus  $p_s / \overline{s} = 1$. 
The ratio $p_s / \overline{s}$ represents a measure of the trade-off between the increase in both the success ratio and the stretch. The larger the ratio, the more efficient the navigation. 

For each network under consideration, it exists an optimal value $a_O$ of the activation probability that maximizes the trade-off between success ratio and stretch. For the ArXiv and the Commodities, the curves of the ratio $p_s / \overline{s}$ as a function of $a$ depend on the duration $T$, with larger $p_s / \overline{s}$ for smaller $T$, while for the Internet and the WTW, these curves are independent of $T$ and collapse. Fig.~\ref{fig:stretch_a} shows that the WTW combines the largest success ratio with the smallest stretch, followed by the Internet, Metabolic, Commodities and the ArXiv networks.

However, it is important to remark that the $p_s/\overline{s}$ ratio is always higher for the static maps than for the temporal ones. 
For instance, the static value for the Internet is $p_s/\overline{s}=0.87$ (see Tab.~\ref{tab:1} in Methods), while in the temporal network it does not exceed $0.70$. 
This is due to the fact that, in temporal maps the large gain in success, which is bounded with a top value of $1$, necessarily comes with an increase in stretch, which in fact can be quite limited, as for the Internet or the WTW, but is unbounded.

\subsection{\label{sec:level2}Heterogeneous activation dynamics}

In this section, we analyse how navigation is affected by an activation probability which varies across nodes. We do it in two different fashions: i) constant activation probability $a<1$ only for nodes whose degree belongs to a certain interval, and  ii) activation probability linearly depending on nodes' degrees.

\subsubsection{Activation of nodes within degree intervals }
\label{sec:level3}

\begin{figure*}[tb]
  \begin{center}
    \includegraphics[width=1.0\textwidth]{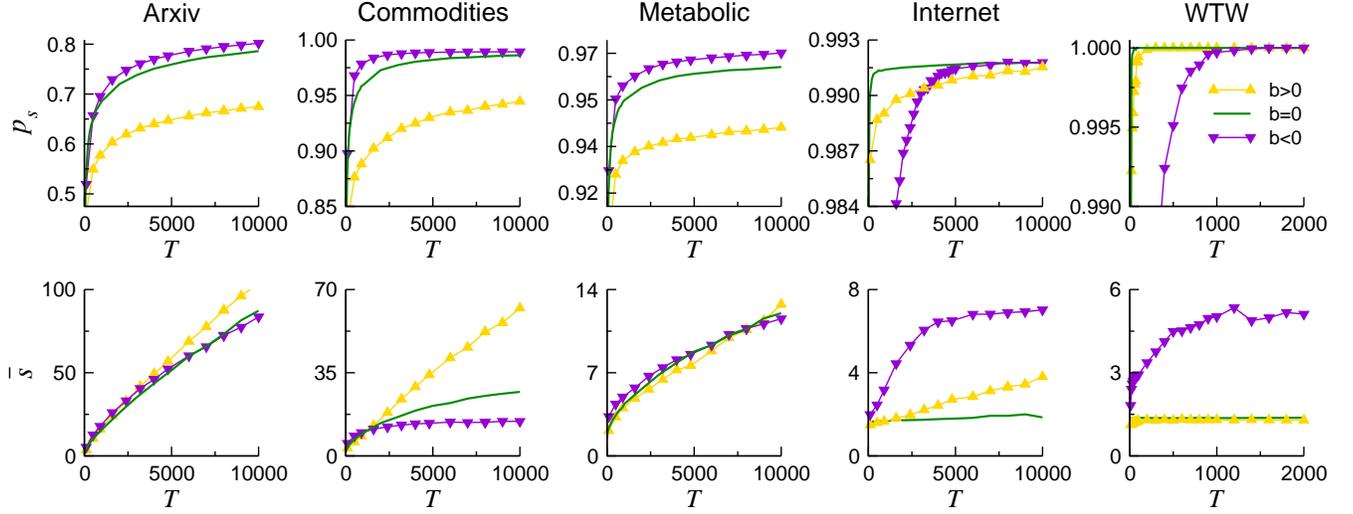}
  \end{center}
  \caption{\textbf{Success ratio $p_s$ (top row) and average stretch $\overline{s}$ (bottom row) as a function of $T$} for $\overline{a}=0.5$, see Fig. S4 in SM for other values of $\overline{a}$. Solid lines designate constant activation of nodes, while symbols indicate that nodes activate linearly and proportionally to their degree ($b>0$), or inversely proportional ($b<0$). The minimum and maximum activation probabilities allowed are $10^{-3}$ and $1-10^{-3}$ respectively.
  As a reference, the values for the static networks, $a=1$, are given in Table~\ref{tab:1} in Methods.}
      \label{fig:Linear}
\end{figure*} 

Here, the random activation dynamics is targeted to subsets of equal number of nodes with degrees in a certain range of values. We measure the success ratio when only one of these subsets of nodes is randomly activated-inactivated with constant $a$ and the rest of the network remains active. To implement this prescription, we order all nodes in a network from highest to lowest degree and divide this sorted list in segments of same number of nodes. The node bins are then labeled using the average degree $\overline{k}$ of the nodes belonging to that bin. The size of the bins has been set to $\xi=5\%$ of the total number of nodes $N$.

This method aims at identifying which degree intervals have a major contribution to the destruction of topological traps, and hence specially boost the success. We find that all temporal maps $\mathcal{M}(\mathcal{G},S)$ experience a sudden increase in $p_s$ when $\overline{k}$ is at its maximum, see Fig. S2 in SM. In scale-free networks, node degrees are distributed as a power-law $p(k)\sim k^{-\gamma}$ (the $\gamma$ values for the considered networks are reported in Table~\ref{tab:1}, Methods). This means that the interval $\overline{k}_{\mathrm{max}}$ contains not only the biggest hub but also several densely connected nodes.  
Therefore, our results imply that switching on and off nodes with low degree has a limited effect on the efficiency of navigation. In fact, the higher the degree of a node the more radical the changes it can induce in the direction of a greedy path.\\

When we construct the bin with temporal behaviour by sampling the nodes uniformly at random from any part of the degree spectrum, so that the bin is approximately characterized by the $\overline{k}$ of the entire network, we find a low $p_s$ close to the static and similar to that obtained for bins of low degree nodes. Therefore, not all nodes are equally able of beating topological traps, and the increase in success mostly relies on activation dynamics affecting densely connected nodes. Also, notice that activating with $a<1$ some randomly selected nodes is not equivalent to activating the whole network with the corresponding average probability $\overline{a}=1-\xi(1-a)$. This is due to the fact that, while navigating the network, some nodes are visited more often than others. If all visited nodes can activate with some $\overline{a} < 1$, the actual noise affecting the network becomes effectively greater than in our implementation.\\

When information packets are able to escape cycles, and the success rises due to the emergence of new (lengthier) successful paths, the average stretch increases too. We corroborate this statement in the bottom row of Fig. S2 in SM, which shows that the maximum stretch precisely occurs at $\overline{k}_{\mathrm{max}}$ in all networks. As expected, the highest $\overline{s}$ is always found for the minimum activation $a=0.1$, which corresponds to the situation where packets find most nodes along their routes to be inactive so they are constantly redirected, thus producing long greedy paths. Interestingly, at $\overline{k}_{\mathrm{max}}$ the $\overline{s}$ for $a=0.9$ considerably varies across networks, with the ArXiv exhibiting the highest value, then Commodities, Metabolic, Internet and finally the WTW. This supports the idea that more congruent networks posess less topological traps. In general it is also satisfied that $a$ values around $a_O$ display  lower $\overline{s}$ at $\overline{k}_{\mathrm{max}}$. 

\subsubsection{\label{sec:level3}Linear activation depending on degree}

Here, we study the navigability of temporal maps in which the activation probability of a node $i$, $a_i$, depends linearly on its degrees $k_i$, such as
\begin{equation}
a(k) = bk +c.
\label{eq:1}
\end{equation}
For $b>0$, the activation probability is proportional to the node's degree so the larger the degree the more active is the node; if $b<0$,  the activation probability is inversely proportional to $k$, and if $b=0$ we recover the case of constant activation probability.
The average activation probability $\overline{a}$ of the whole network, $\overline{a}=N^{-1} \sum^{N}_{i=1} a_{i}(k)$, varies  depending on the choices of the coefficients $b$ and $c$. 
We set the average activation probability $\overline{a}$ as an independent parameter, and choose $c$ so that $c =\overline{a}-b\overline{k}$.
The constraints for the coefficient $b$ arise from the network's minimum and maximum degree, $k_{\min}$ and $k_{\max}$, respectively, as
\begin{eqnarray}
b \leq \mathrm{min} \left\lbrace \frac{1-\overline{a}}{(k_{\mathrm{max}}-\overline{k})}, \frac{-\overline{a}}{(k_{\mathrm{min}}-\overline{k})} \right\rbrace  & \qquad \textrm{if} \quad b>0 \nonumber \\
\qquad |b|\leq \mathrm{min} \left\lbrace \frac{-(1-\overline{a})}{(k_{\mathrm{min}}-\overline{k})}, \frac{\overline{a}}{(k_{\mathrm{max}}-\overline{k})} \right\rbrace  & \qquad \textrm{if} \quad b<0.  \nonumber 
\end{eqnarray}\
For each network under consideration, we choose two values of the coefficient $b$ (one positive and one negative) that ensure the highest heterogeneity in the activation probability, without completely inactivating any of the nodes, i.e.  $a(k_{\min})>0$ and  $a(k_{\max})>0$.
Details regarding the choice of the coefficient can be found in the SM. 

Figure~\ref{fig:Linear} shows the effects of a heterogeneous activation probability of nodes, both proportionally ($b>0$) and inversely proportional ($b<0$) to their degree $k$, compared with constant activation probability ($b=0$), for $\overline{a}=0.5$. 
We observe that when the activation probability is proportional to $k$, 
all $\mathcal{M}(\mathcal{G},S)$ tend to exhibit lower success ratios than in the case with the same $\overline{a}$ but constant activation probability. 
This effect can be understood by considering that highly connected nodes are visited more often than the rest during the routing process. Consequently, the system exhibits an effective $\overline{a}$ higher than $0.5$, which induces 
lower $p_s$ values (closer to the static reference) as shown in previous results. The same reasoning explains the observed behaviour of the stretch $\overline{s}$ in bottom row of Fig.~\ref{fig:Linear}. For all networks and when $b>0$, the tendency of $\overline{s}$ is comparable to that found for $\overline{a}>0.5$ in the constant activation case, see bottom row Fig.~\ref{fig:success}. This feature is specially noticeable for the Internet, where $\overline{s}$ growth is similar to the obtained for $\overline{a}=0.9$.

On the opposite situation, for $b<0$, when the activation probability gets lower as nodes become more connected, the reversed phenomenon occurs.  Figure~\ref{fig:Linear} (bottom row) shows that for all $\mathcal{M}(\mathcal{G},S)$, the $\overline{s}$  resembles that found for low constant $\overline{a}$'s. Moreover, a small increase in $p_s$ with respect to $b=0$ is visible for the ArXiv, the Commodities, and the Metabolic networks in top row of Fig.~\ref{fig:Linear}. The cause is an effective overall activation $\overline{a}<0.5$. For the Internet, $p_s$ still grows higher for $b<0$ than for $b=0$, though it is barely appreciable, and it requires large $T$. For the WTW, $p_s$ reaches $1$ for any $b$ so we can not observe an incremented $p_s$. However, we note that $p_s$ growth for the WTW is retarded when $b<0$, just as happens for the Internet. In these two last networks, the influence of the largest hub on the routing performance is remarkable due to their strong hierarchical nature. For this reason, the time $T$ needed for achieving $p_s\sim 1$ is noticeably enlarged in the event that the main hub is poorly activated.

\section{\label{sec:level1}Discussion}

Navigability is a primary function in many complex networks that, as we have shown, can be strongly affected by temporal alterations in the activity of nodes.
The interplay between the activation dynamics generating the temporal networks and the greedy routing process, indeed, yields a rich phenomenology. The activation process can be understood as the result of random events, like service failures, or, alternatively, it could be thought as part of a local information transfer protocol applied by the node holding the packet in a static network, so to boost the success of the routing operation at a limited cost. 

Our results show that, surprisingly, temporal maps can be navigated more efficiently than the corresponding static ones, even though the number of simultaneously available paths to transfer the packets is greatly reduced.  
 Interestingly, the number of successful paths, in which the packet reaches its destination, is increased due to the activation dynamics. This increase in the success ratio $p_s$ comes at the cost of a growth in the stretch $\bar{s}$, meaning that longer paths are required to successfully deliver the packet. 
However, the ratio between the success and the stretch, $p_s / \bar{s}$, shows a non-trivial behavior as a function of the activation probability, unveiling the existence of an optimal value which maximizes the increase in the success and at the same time minimizes the increase in the stretch. 

More realistic forms of the activation probability, i.e., when the dynamics only affects a subset of nodes or when the activation is correlated with the degree, show similar results. This analysis uncovered the role of highly connected nodes in the routing process, 
which are mainly responsible for the larger success ratio achieved in temporal maps.
Contrary to expectations, our findings suggest that it is possible to improve the routing performance by switching on and off the hubs of the network more often than the rest of the nodes. 
Finally, the navigability of some real networks, like the Internet and the WTW, remains extremely high in the temporal maps. In fact, time-varying effects increase even more the high success rate associated to the static maps, at the cost of a very small increase in the stretch, slowly growing with T, a feature that we name ultranavigability. Even more, temporal changes in the structure of these networks increase the success even if the activation probability is very low.
At the same time, the high routing success observed in these networks could be due in part to temporal behavior in the system, although this possibility has not been acknowledged before and all the merit for their navigability properties has been accredited to their static architecture.

Our work sets a first attempt to measure the effects of temporal dynamics on the navigability of real networks. 
It has been increasingly recognized, indeed, that networks are dynamic entities that evolve in time, with connections being established and terminated for different reasons. 
This study paves the way towards a better understanding of the role of the network's temporal dimension in navigation processes, and provides hints for developing better routing strategies exploiting such dynamics.
Further research is in order to extend our results. 
One may consider more sophisticated generative models of temporal networks,
that may, for instance, incorporate a bursty dynamics of links or nodes.

\section{\label{sec:level1}Methods}
\subsection{Empirical data}
Here we give a brief description of the five networks considered in our study and source references for their data. In Table~\ref{tab:1}, we report the values of different metrics for the five networks. Notice that we used the giant connected component in all cases.

\setlength{\tabcolsep}{8.5pt}
\begin{table}[htbp]
\centering
\renewcommand*{\arraystretch}{1.3}
 \begin{tabular}{cccccccc}
\hline\hline
 \textbf{Network} & $N$ & $E$ & $\overline{k}$ & $k_{\mathrm{max}}$  & $\gamma$ & $p_s$ & $\overline{s}$\\ 
\hline 
ArXiv & $2121$ & $5473$ & $5.16$ & $70$ & $2.86$ & $0.24$ & $1.14$\\ 
\hline
Commodities & $374$ & $1090$ & $5.83$ & $86$ & $2.61$ & $0.57$ & $1.19$\\ 
\hline
Metabolic & $1008$ & $3285$ & $6.51$ & $143$ & $2.53$ & $0.77$ & $1.17$ \\ 
\hline
Internet & $23748$ & $58414$ & $4.92$ & $2778$ & $2.10$ & $0.97$  & $1.11$\\ 
\hline
WTW & $189$ & $550$ & $5.82$ & $110$ & $2.22$ & $0.98$ & $1.04$ \\
\hline\hline
\end{tabular}
\caption{\small{\textbf{Topological properties and navigation performance values of five real static maps $\mathcal{M}(G_0,S)$.} From left to right: number of nodes, number of edges, average degree, maximum degree, exponent of the power-law degree distribution, success ratio and average topological stretch.}}
\label{tab:1}
\end{table}

\noindent\textbf{ArXiv.} The ArXiv network is a graph representing co-authorship of papers~\cite{Manlio2015}, elaborated from data of the free scientific repository ArXiv. The nodes are authors which are connected if they have co-authored a paper belonging to category “Disordered Systems and Neural Networks” (cond-mat.disnn). The data considers only papers from up to May 2014, with the word “networks” in the title or abstract. The hyperbolic embedding of this network comes from Ref.~\cite{Kleineberg2016}.\\
\textbf{US Commodities.} The commodities network~\cite{Grady2012} is a representation of the flows of services and goods (in USD) exchanged between industrial sectors during year 2007 in the United States. The hyperbolic embedding of this network comes from Ref.~\cite{allard2017the}.\\
\textbf{Metabolic.} The metabolic network used in this study corresponds to the one-mode projection of metabolites of the bipartite metabolic network of the bacterium \textit{E. coli}. In this representation, two metabolites are connected if they participate in the same biochemical reaction. We use the data originally extracted from the BiGG database \cite{BiGG} and reconstructed in~\cite{Serrano2011} as a spatially embedded network.\\
\textbf{Internet.} We use the connectivity data for the Internet at the autonomous systems level collected by the Archipelago project \cite{ClaffyK} corresponding to June 2009 and reconstructed as a network embedded in hyperbolic space in Ref.~\cite{Boguna:2010aa}.\\ 
\textbf{WTW.} The world trade web consists of significant trade exchanges between countries. We use the network corresponding to the most recent data available, the year 2013, and its embedding as given in Ref.~\cite{garciaperez2016}.\\

\subsection{Hyperbolic maps}
In the hyperbolic hidden metric space, every node $i$ has coordinates ($r_i$,$\theta_i$).
Embedding a network in a metric space means constructing a map that correlates topology and geometry. 
The idea is that a link between two nodes in the topology exists with a certain probability $p(d)$ 
that depends on their distance $d$, measured in the hidden metric space,
 such that nodes with higher probabilities of being connected are closely positioned in the geometric space~\cite{Serrano:2008ga}. 
Therefore, $p(d)$ needs to be a decreasing function of distance between nodes. In the hyperbolic plane, the distance between nodes $d_{ij}$ is calculated using the hyperbolic law of cosines:
\begin{equation}
\mathrm{cosh}(d_{ij})=\mathrm{cosh}r_i\mathrm{cosh}r_j - \mathrm{sinh}r_i\mathrm{sinh}r_j\mathrm{cos}\Delta\theta_{ij}
\end{equation}

It has repeatedly been shown that real networks can be embedded into a hyperbolic plane \cite{Boguna:2010aa,Serrano2011} ,
in a way that all relevant topological properties of the networks, 
such as the small world property, the degree distribution, degree-degree correlations, the clustering coefficient, and degree-thresholding topological self-similarity~\cite{Serrano:2008ga} are reproduced by the model.


\section{\label{sec:level1}Acknowledgements}

This work was supported by a James S. McDonnell Foundation Scholar Award in Complex Systems, the Ministerio de Econom\'{\i}a y Competitividad of Spain projects no. FIS2013-47282-C2-1-P and no. FIS2016-76830-C2-2-P (AEI/FEDER, UE), and the Generalitat de Catalunya grant no. 2014SGR608.

\section{\label{sec:level1}Author contributions statement}
E.O., M.S. and M.\'{A}.S. contributed to the design and implementation of the research, to the analysis of the results, and to the writing of the manuscript.

\section{\label{sec:level1}Additional information}
\textbf{Competing financial interests:} The authors declare no competing financial interests.

\end{document}


\beginsupplement

\author{Elisenda Ortiz Castillo}
\author{Michele Starnini}
\affiliation{Departament de F\'{\i}sica de la Mat\`eria Condensada, Universitat de Barcelona, Mart\'{\i} i Franqu\`es 1, 08028 Barcelona, Spain}
\affiliation{Universitat de Barcelona Institute of Complex Systems (UBICS), Universitat de Barcelona, Barcelona, Spain}
\author{M. Ángeles Serrano}
\email{marian.serrano@ub.edu}
\affiliation{Departament de F\'{\i}sica de la Mat\`eria Condensada, Universitat de Barcelona, Mart\'{\i} i Franqu\`es 1, 08028 Barcelona, Spain}
\affiliation{Universitat de Barcelona Institute of Complex Systems (UBICS), Universitat de Barcelona, Barcelona, Spain}
\affiliation{ICREA, Pg. Llu\'{\i}s Companys 23, E-08010 Barcelona, Spain}
\date{\today}

\title{Supplementary information for:\\ Navigability of temporal networks in hyperbolic space}

\maketitle

\section{Greedy routing in hyperbolic temporal maps}
\begin{figure}[H]
\centering
    \includegraphics[width=0.60\textwidth]{\FigPath/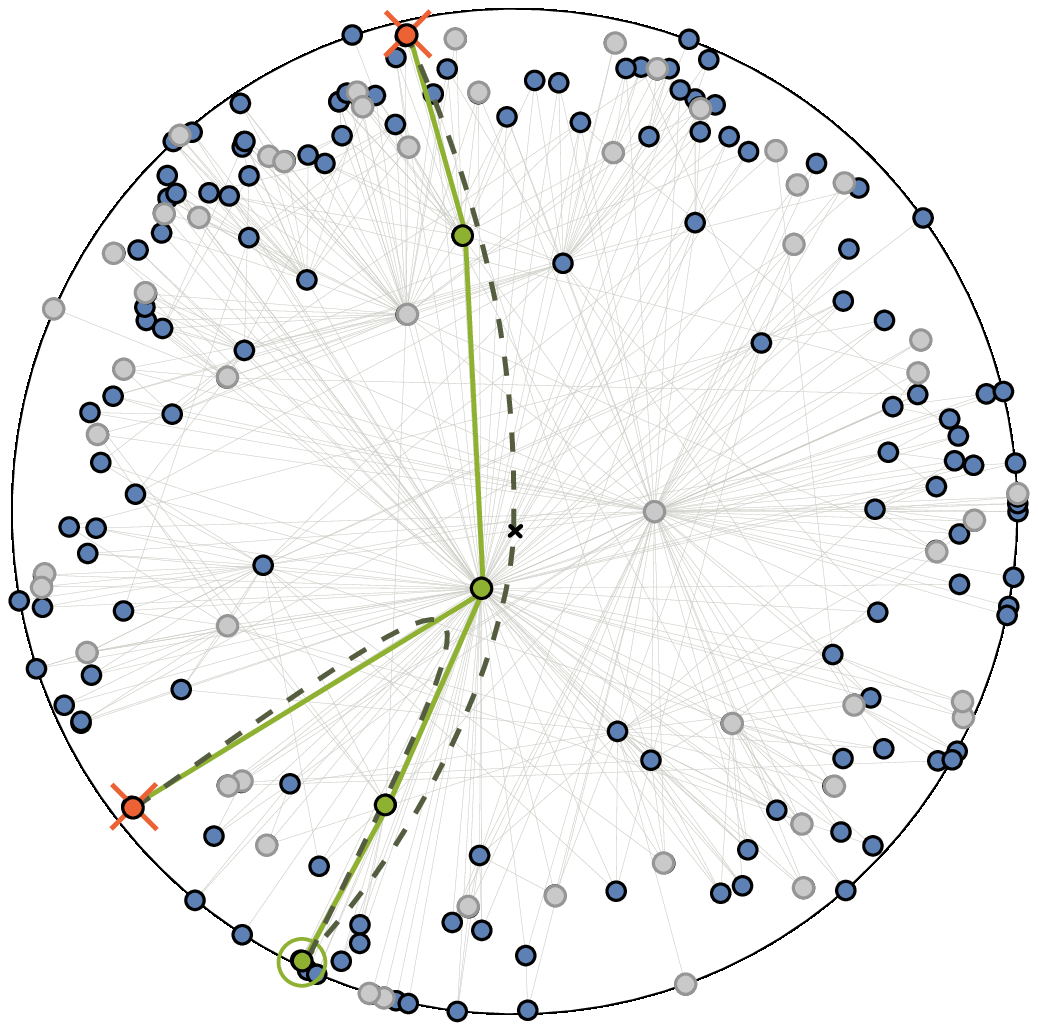}
  \caption{\textbf{Visualization of the WTW temporal map $\mathcal{M}(\mathcal{G},S)$ in the hyperbolic plane and greedy routig in it.} Inactive nodes at time $t$ appear in grey while blue color is reserved for active ones. The two dashed lines correspond to two geodesics between the source node, circled in green at the bottom, and two destinations represented by red nodes with crosses. The solid green lines, which are the greedy paths between the same source-destination nodes, are remarkably congruent with the geodesics. In this example, if a green node along a greedy path (specially the most centric one) is switched off at the time of receiving a packet, it will force the greedy path to deviate from the geodesic and turn it into a longer greedy path.}
      \label{fig:Hyperbolic_navigation}
\end{figure}

\newpage
\section{Activation of nodes within degree intervals}
\begin{figure}[htbp]
    \includegraphics[width=1.0\textwidth]{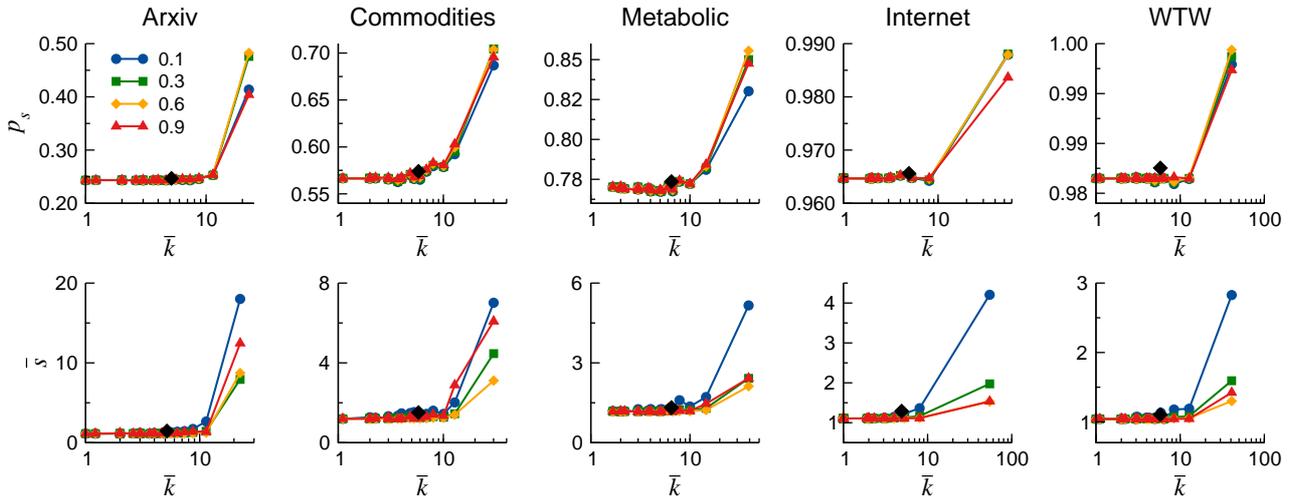}
  \caption{\textbf{Success ratio $p_s$ (top row) and average stretch $\overline{s}$ (bottom row) as a function of the average degree $\overline{k}$ of node bins} with temporal behaviour, for different values of the activation probability $a$. The success ratio is measured in a fully active network, where only nodes from a single bin of size $\xi=5\%$ can activate with probability $a < 1$.  The black diamonds  correspond to the case of selecting the number of nodes that activate with $a<1$ uniformly at random from the whole network and averaging over $10^3$ realizations. In this later case, the results for different $a$'s overlap, so for clarity the black diamonds are displayed only for $a=0.1$.}
      \label{fig:degree_sections}
\end{figure}

In this section the random activation-inactivation dynamics is targeted to subsets of equal number of nodes whose degrees belong to a certain degree interval. Figure. \ref{fig:degree_sections} shows the $p_s$ and $\overline{s}$, measured when only one of these subsets of nodes is activated with constant $a <1$ and the rest of the network remains active ($a=1$). The dynamic subsets are identified by their average degree $\overline{k}$ and are of size $\xi=5\%$ of the total number of nodes $N$. Results from Fig. \ref{fig:degree_sections} imply that switching on and off nodes with low degree has a limited effect on the navigability efficiency. Moreover, when the dynamic subset is composed by nodes of any degree, the obtained $p_s$ remains low and close to the static (see black diamonds, Fig. \ref{fig:degree_sections}) demonstrating that the incresed success achived by temporal networks relies on the activity of densely connected nodes.

\section{Linear activation dynamics of nodes}\label{sec:Linear_activation}
In this section, we set the activation probabilities to be linearly dependent upon the degree ($k$) of the nodes. This is, $a(k) \in [0,1]$ takes the form:
\begin{equation}
a(k) = bk +c
\label{eq:1}
\end{equation}

We tune the average activation probability $ \overline{a}=\frac{1}{N}\sum^{N}_{i=1} a(k)_{i}$ of the whole network, so for each $\overline{a}$ we calculate the corresponding coefficients $b$ and $c$ taking into account that $\overline{a}$ is set constant. Then, we activate each node proportionally to its degree following \eqref{eq:1}. The coefficients $b$ and $c$ are obtained as follows.
 
Given a fixed $\overline{a}$ value, we have that
\begin{equation}
 \label{eq:2}
  \begin{gathered}[b]
    \overline{a} = \frac{1}{N} \sum_{i=1}^{N}{(bk_i + c)}\\
    c =\overline{a}-\frac{b}{N}\sum_{i=1}^{N}k_i\\
    c =\overline{a}-b\overline{k}
  \end{gathered}
\end{equation}

\begin{figure}[htbp]
  \begin{center}
    \includegraphics[width=0.65\textwidth]{\FigPath/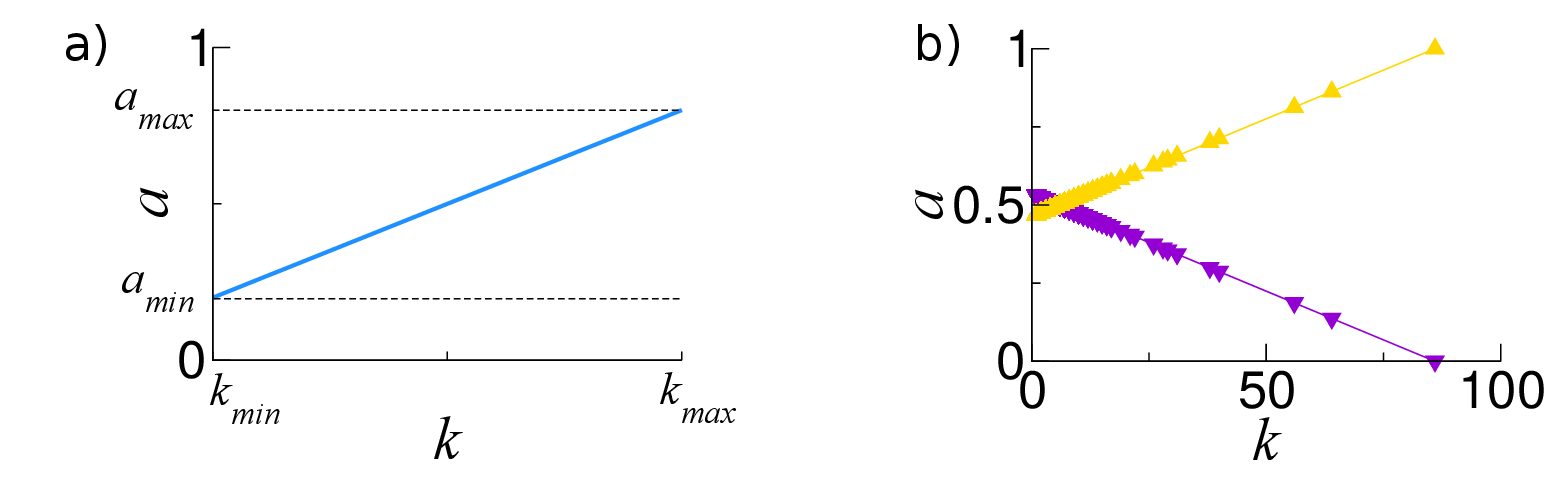}
  \end{center}
  \caption{\textbf{a)} In blue, example of a linear activation probability. \textbf{b)} Example, from Commodities network, of the two curves $a(k)$, used for linearly activating the nodes when $\overline{a}=0.5$. In yellow $a(k)$ with $b>0$ and in purple with $b<0$. The minimum activation probability is explicitly set to not to be exactly 0 but of order $10^{-3}$; simetrically the maximum activation corresponds to $a\approx 1-10^{-3}$.}
      \label{fig:example_linear}
\end{figure}

The maximum ($a_{\mathrm{max}}$) and minimum ($a_{\mathrm{min}}$) values of $a(k)$ are related with $k_{\mathrm{max}}$ and $k_{\mathrm{min}}$ through the equation of a line, see Fig. \ref{fig:example_linear}a. Moreover, recall that $k_{\mathrm{min}}$ , $k_{\mathrm{max}}$ are fixed numbers since they are attributes of the network, and $a_{\mathrm{min}}$, $a_{\mathrm{max}}$ $\in $ $[0,1]$ because $a(k)$ is a probability. Therefore, it is satisfied that

\noindent\begin{minipage}{.45\linewidth}
\begin{equation*}
\begin{aligned}[c]
&\text{$\LargerCdot$ \underline{$b>0$:}} &\qquad a_{\mathrm{max}}&=bk_{\max}+c\\
&&&=bk_{\mathrm{max}}+\overline{a}-b\overline{k}\\
&&&=\overline{a}+b(k_{\mathrm{max}}-\overline{k})
\end{aligned}
\end{equation*}
\end{minipage}%
\begin{minipage}{.45\linewidth}
\begin{equation*}
\begin{aligned}[c]
a_{\mathrm{min}}&=bk_{\min}+c\\
&=\overline{a}+b(k_{\mathrm{min}}-\overline{k})
\end{aligned}
\end{equation*}\\
\end{minipage}

\noindent\begin{minipage}{.45\linewidth}
\begin{equation*}
\begin{aligned}[c]
&\text{$\LargerCdot$ \underline{$b<0$:}} &\qquad a_{\mathrm{max}}&=\overline{a}-|b|(k_{\mathrm{min}}-\overline{k})\\
\end{aligned}
\end{equation*}\
\end{minipage}%
\begin{minipage}{.45\linewidth}
\begin{equation*}
\begin{aligned}[c]
a_{\mathrm{min}}&=\overline{a}-|b|(k_{\mathrm{max}}-\overline{k})\\
\end{aligned}
\end{equation*}\
\end{minipage}

If we now apply $a_{\mathrm{max}}\leq 1$ and $a_{\mathrm{min}}\geq 0$ in order to ensure that $a(k) \in [0,1]$,

\noindent\begin{minipage}{.45\linewidth}
\begin{equation*}
\begin{aligned}[c]
&\text{$\LargerCdot$ \underline{$b>0$:}} &\qquad \qquad \overline{a} + b(k_{\mathrm{max}}-\overline{k}) &\leq 1\\
&&b \leq \frac{1-\overline{a}}{(k_{\mathrm{max}}-\overline{k})}\\
\end{aligned}
\end{equation*}
\end{minipage}%
\begin{minipage}{.45\linewidth}
\begin{equation*}\
\begin{aligned}[c]
&\overline{a} + b(k_{\mathrm{min}}-\overline{k})\geq 0\\
&b\leq -\frac{\overline{a}}{(k_{\mathrm{min}}-\overline{k})}\\
\end{aligned}
\end{equation*}
\end{minipage}

\noindent\begin{minipage}{.45\linewidth}
\begin{equation*}
\begin{aligned}[c]
&\text{$\LargerCdot$ \underline{$b<0$:}} &\qquad \qquad \overline{a} - |b|(k_{\mathrm{min}}-\overline{k}) &\leq 1\\
&&|b| \leq -\frac{(1-\overline{a})}{(k_{\mathrm{min}}-\overline{k})}
\end{aligned}
\end{equation*}\
\end{minipage}%
\begin{minipage}{.45\linewidth}
\begin{equation*}\
\begin{aligned}[c]
&\overline{a} - |b|(k_{\mathrm{max}}-\overline{k})\geq 0\\
&|b|\leq \frac{\overline{a}}{(k_{\mathrm{max}}-\overline{k})}\\
\end{aligned}
\end{equation*}\
\end{minipage}

From the above conditions we obtain that $b$ must satisfy

\begin{eqnarray}
b \leq \mathrm{min} \left\lbrace \frac{1-\overline{a}}{(k_{\mathrm{max}}-\overline{k})}, \frac{-\overline{a}}{(k_{\mathrm{min}}-\overline{k})} \right\rbrace  & \qquad \textrm{if} \quad b>0 \label{eq:3} \\
\qquad |b|\leq \mathrm{min} \left\lbrace \frac{-(1-\overline{a})}{(k_{\mathrm{min}}-\overline{k})}, \frac{\overline{a}}{(k_{\mathrm{max}}-\overline{k})} \right\rbrace  & \qquad \textrm{if} \quad b<0.  \label{eq:4}
\end{eqnarray}\

\begin{figure}[htbp]
\includegraphics[width=160mm]{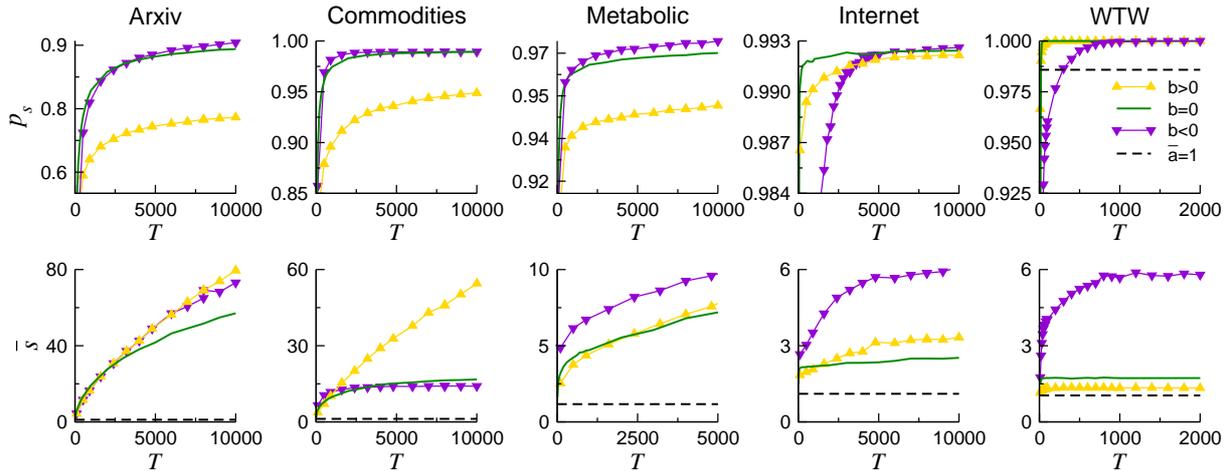} 
\caption{\textbf{Success ratio $p_s$ (top row) and average stretch $\overline{s}$ (bottom row) as a function of $T$}, for $\overline{a}=0.3$, in five temporal maps $\mathcal{M}(\mathcal{G},S)$. Solid lines designate constant activation of nodes, while symbols indicate nodes linearly activate following \eqref{eq:1}; proportionally to their $k$ ($b>0$), or inversely proportional ($b<0$). In dashed line, $p_s$ and $\overline{s}$ corresponding to greedy routing in the static maps $\mathcal{M}(G_0,S)$.}
\label{fig:Fig2}
\end{figure}

\begin{figure}[htbp]
\includegraphics[width=160mm]{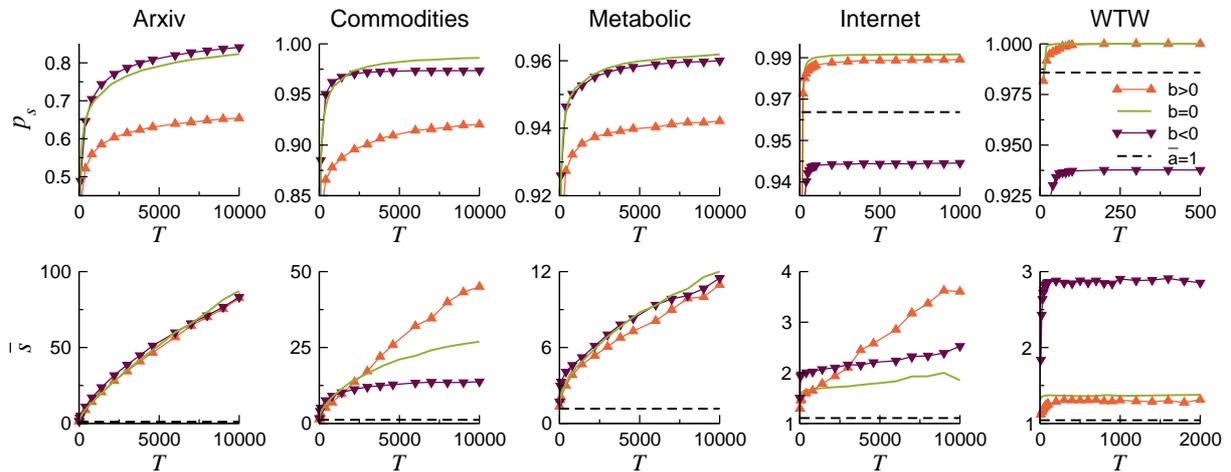} 
\caption{\textbf{Success ratio $p_s$ (top row) and average stretch $\overline{s}$ (bottom row) as a function of $T$}, for $\overline{a}=0.5$, in five temporal maps $\mathcal{M}(\mathcal{G},S)$. In this case, the minimum activation of main hub is allowed to be null and the maximum exactly $1$. Solid lines designate constant activation of nodes, while symbols indicate nodes linearly activate following \eqref{eq:1}; proportionally to their $k$ ($b>0$), or inversely proportional ($b<0$). In dashed line, $p_s$ and $\overline{s}$ corresponding to greedy routing in static maps $\mathcal{M}(G_0,S)$. Notice the low $p_s$ and $\overline{s}$ achieved by the Internet and the WTW  when $b<0$, due to the main hub being always totally inactive during the routing process.}
\label{fig:Fig3}
\end{figure}

Therefore, we can select any $b$ values that satisfy the above inequalities. Afterwards,  the coefficient $c$ is readily found from Eq. \ref{eq:2}. 
We take one positive and one negative $b$, 
so to maximize the heterogeneity of the activation dynamics, as shown in Fig. \ref{fig:example_linear}b. By choosing $b$ that correspond to the equality sign in (3) and (4) we fully activate ($b>0$) or inactivate ($b<0$) the main hub. For this particular case, one can observe the resulting $p_s$ and $\overline{s}$ in Fig.~\ref{fig:Fig3}. Since we noticed that strongly hierarchical networks such as the Internet and the WTW present a singular behaviour when the main hub is removed from the routing ($b<0$), 
we exclude this choice of coefficient and set the minimum activation probability for the main hub to be $10^{-3}$, and simetrically the maximum to $1-10^{-3}$. Figure \ref{fig:Fig2} shows the results of linear activation of nodes for a an average activation probability of $\overline{a}=0.3$, different from the one employed in the main text ($\overline{a}=0.5$). In this case, all temporal maps exhibit the same trends as before, both for $p_s$ and $\overline{s}$, implying that qualitatively, the effects induced by linear activation are independent of $\overline{a}$.

\section{Average geometric stretch}
The average geometric stretch $\overline{s}_g$ of successful greedy paths, is defined as the ratio between the cummulative hyperbolic distance from the hop-length of greedy paths and the corresponding geodesics. The $\overline{s}_g$ tells us how much the successful greedy paths elongate with respect to the shortest ones in the metric space.

\setlength{\tabcolsep}{10pt}
\begin{table}[htbp]
\centering
\renewcommand*{\arraystretch}{1.3}
 \begin{tabular}{cccccc}
\hline\hline
  & ArXiv & Commodities & Metabolic & Internet & WTW \\ 
\hline 
$\overline{s}_g$ & $3.10$ & $2.33$ & $2.38$ & $2.39$ & $1.35$ \\
\hline\hline
\end{tabular}
\caption{\textbf{Average geometric stretch of five static maps $\mathcal{M}(G_0,S)$.} The values were obtained averaging over a number of random source--destination pairs, that is the minimum between $10^5$ and $N(N-1)/2$, where $N$ is the number of nodes of the network.}
\label{tab:1}
\end{table}
An inescapable consequence of embedding the network topologies in the hyperbolic space is the fact that very often there exist nodes which share the precise same coordinates. These nodes that occupy the exact same position in the hyperbolic disk were not considered as possible source--destination pairs, since the geodesic connecting them is of length $0$, thus making $\overline{s}_g$ to diverge. The average geometric stretch for temporal maps $\mathcal{M}(\mathcal{G},S)$ was computed
taking only successful paths from a representative subset of randomly chosen node pairs. For each network the number of source--destination pairs was set to be the lowest number between $N(N-1)/2$ and $10^5$ , and was changed at each different $T$.
The behaviour observed for $\overline{s}_g$ is qualitatively similar to the observed for the topological average stretch in Fig.1, in the main paper. This means, \textit{i)} the geometric lengths of greedy paths elongate more for temporal networks with lower static success, due to the breaking of a greater number of topological traps \textit{ii)} the closest the activation probability is to the optimal, $a_O$, the less the greedy paths elongate with respect to geodesics.

\begin{figure}[htbp]
\includegraphics[width=160mm]{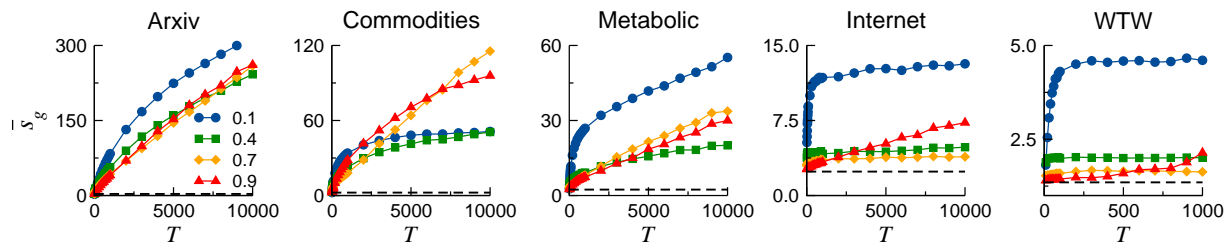} 
\caption{\textbf{Average geometric stretch $\overline{s}$ as a function of $T$}, for different values of the activation probability $a$, in five temporal maps $\mathcal{M}(\mathcal{G},S)$. Average geometric stretch for $a=1$, corresponding to greedy routing in the static maps, is plotted in dashed line. }
\label{fig:geometric_stretch}
\end{figure}

\section{Coverage}
\begin{figure}[htbp]
\includegraphics[width=160mm]{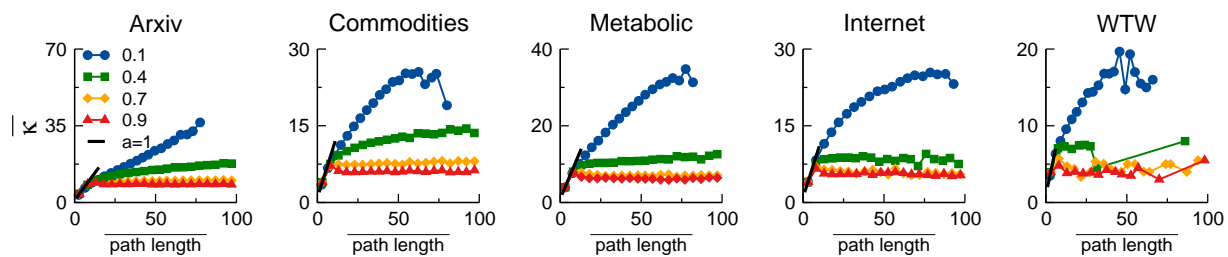} 
\caption{\textbf{Average coverage $\overline{\kappa}$ as a function of the average path length of successful paths} for different values of the activation probability in five temporal maps $\mathcal{M}(\mathcal{G},S)$.  The black solid lines correspond to $a=1$, which indicates the greedy routing is performed on static maps.}
\label{fig:Coverage}
\end{figure}

The average coverage $\overline{\kappa}$, tells the average number of \textit{different} visited nodes, against the average number of nodes that constitute a successful path. This metric measures how much the information packet travels around the network (and the metric space). Figure \ref{fig:Coverage} shows that for low activation probability ($a=0.1$), the information packet jumps among a great number of different nodes, whereas for high activation ($a=0.9$) it only visits a few. By lowering the activity of the nodes, the information packet is forced to jump often to any available position, hence we find that the longer the path the higher the coverage. On the contrary, when most nodes remain almost invariably active, the longer path lengths displaying low $\overline{\kappa}$ confirm the idea that the packet is found moving in cycles, failing to scape a topological trap.
For the static references, plotted as solid lines in Fig. \ref{fig:Coverage}, we observe straight lines of slope $1$. In the static case, a path is always declared unsuccessful if a node is tried to be visited twice. Therefore, the average path length always coincides with the number of different visited nodes, thus producing unitary slopes. Nonetheless, the solid lines inform us about the maximum average path length that one can expect from a network only due to its own structure. 

 \bibliography{References.bib}